# Collision and Annihilation of Relative Equilibrium Points Around Asteroids with a Changing Parameter

Yu Jiang[1, 2], Hexi Baoyin[1], Hengnian Li[2]

1. School of Aerospace Engineering, Tsinghua University, Beijing 100084, China
2. State Key Laboratory of Astronautic Dynamics, Xi'an Satellite Control Center, Xi'an 710043, China

Y. Jiang (✉) e-mail: jiangyu_xian_china@163.com (corresponding author)

**Abstract**. In this work, we investigate the bifurcations of relative equilibria in the gravitational potential of asteroids. A theorem concerning a conserved quantity, which is about the eigenvalues and number of relative equilibria, is presented and proved. The conserved quantity can restrict the number of non-degenerate equilibria in the gravitational potential of an asteroid. It is concluded that the number of non-degenerate equilibria in the gravitational field of an asteroid varies in pairs and is an odd number. In addition, the conserved quantity can also restrict the kinds of bifurcations of relative equilibria in the gravitational potential of an asteroid when the parameter varies. Furthermore, studies have shown that there exist transcritical bifurcations, quasi-transcritical bifurcations, saddle-node bifurcations, saddle-saddle bifurcations, binary saddle-node bifurcations, supercritical pitchfork bifurcations, and subcritical pitchfork bifurcations for the relative equilibria in the gravitational potential of asteroids. It is found that for the asteroid 216 Kleopatra, when the rotation period varies as a parameter, the number of relative equilibria changes from 7 to 5 to 3 to 1, and the bifurcations for the relative equilibria are saddle-node bifurcations and saddle-saddle bifurcations.

**Key words**: methods: analytical-planets and satellites: dynamical evolution and stability-minor planets, asteroids, general

## 1. Introduction

The Yarkovsky-O'Keefe-Radzievskii-Paddack(YORP) effect (Chesley et al. 2003; Taylor et al. 2007; Lupishko et al. 2014), the collisions and gravitational reaccumulation (Michel et al. 2001), the surface grain motion (Richardson et al. 2004; Campins et al. 2010), and the disruption and breakup (Asphaug et al. 1998; Nesvorný et al. 2004) cause an asteroid's gravitational field to vary.

The YORP effect can cause the asteroid's rotation speed to increase or decrease depending on the orbit, the irregular shape, and the rotation direction of the asteroid (Rubincam 2000). For instance, it makes the rotation speed of asteroid 54509 (2000 PH5) increase with an accelerated speed of $(2.0 \pm 0.2) \times 10^{-4}$ deg/day$^2$ (Taylor et al. 2007) and that of asteroid 1620 Geographos increase with an accelerated speed of $1.15 \times 10^{-8}$ rad/day$^2$ (Ďurech et al. 2008). The YORP effect can change an asteroid's rotation speed over millions of years, and may cause rotational breakup and produce binary asteroids (Nesvorný et al. 2002; Walsh et al. 2008).

The numerical experiments show that the YORP effect can lead to the disruption of rubble-pile asteroids and the formation of asteroidal moonlets (Walsh et al. 2012). The YORP effect changes the characteristics of equilibrium points and equilibrium shapes, both of which have an influence on the tensile strength and structure of asteroids (Richardson et al. 2005; Hirabayashi and Scheeres 2014).

To simulate the motion around equilibrium points and to aid in the understanding of this motion, some simple-shaped bodies have been considered. Elipe and Lara (2004) used a massive straight segment to model the gravitational field of asteroid 433 Eros and calculated the location of the equilibrium points. Placián et al. (2006) discussed the location and motion around collinear equilibria in the gravitational field of a straight segment. Liu et al. (2011) have investigated the location, linear stability, and periodic orbits around equilibria near a rotating homogeneous cube. Romanov and Doedel (2012) computed several periodic orbit families generated around equilibrium points in the gravitational field of the rotating homogeneous triaxial ellipsoid. Li et al. (2013) studied the location and linear stability of equilibria around a rotating homogeneous dumbbell-shaped body. Using the observation data of asteroids, the irregular shapes and gravitational field of the asteroids can be calculated more precisely, and the location and linear stability of the equilibrium points in the potential of the asteroids can be calculated accurately. Yu and Baoyin (2012) found 4 equilibria outside the body of asteroid 216 Kleopatra and calculated the position of the equilibria and analyzed their linear stability. Jiang et al. (2014) classified the equilibrium points around asteroids using the topological characteristics and applied these classifications to 4 asteroids. Wang et al. (2014) calculated the location and topological cases of several asteroids, comets, and planetary satellites. Hirabayashi and Scheeres (2014) also calculated the position and linear stability of the equilibria of asteroid 216 Kleopatra and compared the result with those of Yu and Baoyin (2012). Jiang (2015) discussed the correspondence of topological types of periodic orbits around equilibrium points and the topological types of equilibrium points. When the parameter of the asteroids varies, the location and stability may change (Romanov and Doedel 2012; Yu and Baoyin 2012; Hirabayashi and Scheeres 2014; Jiang et al. 2015). The parameter may be the rotation speed, the shape, the density, etc. and can be changed by the YORP effect (Taylor et al. 2007), the collisions (Michel et al. 2001), the surface grain motion (Richardson et al. 2004), etc.

We consider the rotation speed as a parameter in order to calculate the change of the relative equilibria in the gravitational field of an asteroid. The number of relative equilibria will vary and the relative equilibria will collide and annihilate each other. We presented and proved a theorem concerning the eigenvalues and the number of relative equilibria. Using this theorem, it is discovered that the number of non-degenerate equilibria in the potential of an asteroid varies in pairs and is an odd number. Some kinds of bifurcations are likely to appear when the parameter varies, including transcritical bifurcations, quasi-transcritical bifurcations, saddle-node bifurcations, saddle-saddle bifurcations, binary saddle-node bifurcations, supercritical pitchfork bifurcations, and subcritical pitchfork bifurcations.

Asteroid 216 Kleopatra, which has an incompact rubble-pile structure (Ostro et al. 2000), is taken as an example, when the parameter varies, the number of relative equilibria changes from 7 to 5 to 3 to 1. When the parameter varies, the equilibrium points E3 and E6 (see Sect. 3 for a definition of E1 through E7 around asteroid 216 Kleopatra) collide and annihilate each other; the bifurcation produced is a saddle-node bifurcation and occurs at $\omega = 1.944586\omega_0$, where $\omega$ is the rotation speed and $\omega_0$ is the initial rotation speed. After the annihilation, there are only 5 equilibrium points left. When the parameter varies to $\omega = 2.03694\omega_0$, the equilibrium points E1 and E5 collide and annihilate each other; the bifurcation produced is also a saddle-node bifurcation. After the annihilation, there are only 3 equilibrium points left. When the parameter varies to $\omega = 4.270772\omega_0$, the equilibrium points E4 and E7 collide and annihilate each other; the bifurcation produced is a saddle-saddle bifurcation. After the annihilation, there is only 1 equilibrium point left.

**2. Gravitational Field, Effective Potential, and Relative Equilibria**

Consider the motion of a massless particle in the potential of an asteroid. The asteroid's gravitational potential can be computed by the polyhedron method (Werner and Scheeres 1997)

$$U = \frac{1}{2}G\sigma \sum_{e \in edges} \mathbf{r}_e \cdot \mathbf{E}_e \cdot \mathbf{r}_e \cdot L_e - \frac{1}{2}G\sigma \sum_{f \in faces} \mathbf{r}_f \cdot \mathbf{F}_f \cdot \mathbf{r}_f \cdot \omega_f. \quad (1)$$

Where $G = 6.67 \times 10^{-11}$ m$^3$kg$^{-1}$s$^{-2}$ represents the asteroid's gravitational constant, $\sigma$ represents the bulk density of the asteroid; $\mathbf{r}_e$ and $\mathbf{r}_f$ are vectors from some fixed points to points on the edge $e$ and face $f$, respectively; $\mathbf{E}_e$ and $\mathbf{F}_f$ are tensors of edges and faces, respectively; $L_e$ is the integration factor, and $\omega_f$ is the signed solid angle. The gradient of the gravitational potential (Werner and Scheeres 1997) is

$$\nabla U = -G\sigma \sum_{e \in edges} \mathbf{E}_e \cdot \mathbf{r}_e \cdot L_e + G\sigma \sum_{f \in faces} \mathbf{F}_f \cdot \mathbf{r}_f \cdot \omega_f. \quad (2)$$

The particle's effective potential is given by (Scheeres et al. 1996; Scheeres 2012; Jiang and Baoyin 2014; Jiang et al. 2015)

$$V(\mathbf{r}) = -\frac{1}{2}(\boldsymbol{\omega} \times \mathbf{r})(\boldsymbol{\omega} \times \mathbf{r}) + U(\mathbf{r}), \quad (3)$$

where $\mathbf{r} = (x, y, z)^{\mathrm{T}}$ is the relative position of the particle in the body-fixed frame of the asteroid, and $\boldsymbol{\omega}$ is the asteroid's rotational angular velocity relative to the inertial frame. The relative equilibrium points in the gravitational field of the asteroid are the solution of the equation $\nabla V(x, y, z) = 0$, where $\nabla$ is the gradient operator.

Eigenvalues of the equilibria have the form $\pm \alpha (\alpha \in \mathrm{R}, \alpha \geq 0)$, $\pm i\beta (\beta \in \mathrm{R}, \beta > 0)$, and $\pm \sigma \pm i\tau (\sigma, \tau \in \mathrm{R}; \sigma, \tau > 0)$. The non-degenerate and non-resonant relative equilibria (Jiang et al. 2014; Wang et al. 2014; Jiang 2015) have five topological cases, they are Case 1: $\pm i\beta_j (\beta_j \in \mathrm{R}, \beta_j > 0; j = 1, 2, 3)$; Case 2: $\pm \alpha_j (\alpha_j \in \mathrm{R}, \alpha_j > 0, j = 1)$ and $\pm i\beta_j (\beta_j \in \mathrm{R}, \beta_j > 0; j = 1, 2)$ ; Case 3:

$\pm \alpha_j \left( \alpha_j \in \mathrm{R}, \alpha_j > 0; j = 1, 2 \right)$ and $\pm i\beta_j \left( \beta_j \in \mathrm{R}, \beta_j > 0, j = 1 \right)$ ; Case 4a: $\pm \alpha_j \left( \alpha_j \in \mathrm{R}, \alpha_j > 0, j = 1 \right)$ and $\pm \sigma \pm i\tau \left( \sigma, \tau \in \mathrm{R}; \sigma, \tau > 0 \right)$ ; Case 4b: $\pm \alpha_j \left( \alpha_j \in \mathrm{R}, \alpha_j > 0, j = 1, 2, 3 \right)$ ; and Case 5: $\pm i\beta_j \left( \beta_j \in \mathrm{R}, \beta_j > 0, j = 1 \right)$ and $\pm \sigma \pm i\tau \left( \sigma, \tau \in \mathrm{R}; \sigma, \tau > 0 \right)$. The topological classification of equilibria is based on the distribution of eigenvalues of equilibria on the complex plane. The non-degenerate and non-resonant equilibrium case 1 has 3 periodic orbit families; while the non-degenerate and non-resonant equilibrium case 2 has 2 periodic orbit families, and the non-degenerate and non-resonant equilibrium cases 3 and 5 have only 1 periodic orbit family each. The degenerate equilibrium case 4 has no periodic orbit family. The degenerate relative equilibria have at least two eigenvalues equal to 0. Figure 1 presents the topological classification of the degenerate equilibrium points. The degenerate equilibrium case D1 has 2 periodic orbit families; while the degenerate equilibrium cases D2 and D5 have only 1 periodic orbit family each. The degenerate equilibrium cases D3, D4, D6, and D7 have no periodic orbit families. Wang et al. (2014) calculated the topological classification of relative equilibria in the gravitational field of 23 minor celestial bodies; which included 15 asteroids, 5 satellites of planets, and 3 comets; all of these equilibria belong to case 1, 2, or 5. They found that for some of the minor bodies, including asteroids 243 Ida, 101955 Bennu, satellites of planets J5 Amalthea, comet 103P/Hartley 2, etc., the topological classification of outside equilibria only belong to case 2 or 5; while for others, including asteroids 4 Vesta, 2867 Steins, satellites of planets N8 Proteus, S9 Phoebe, comets 1P/Halley, etc., the topological classification of outside equilibria only belong to case 1 or 2.

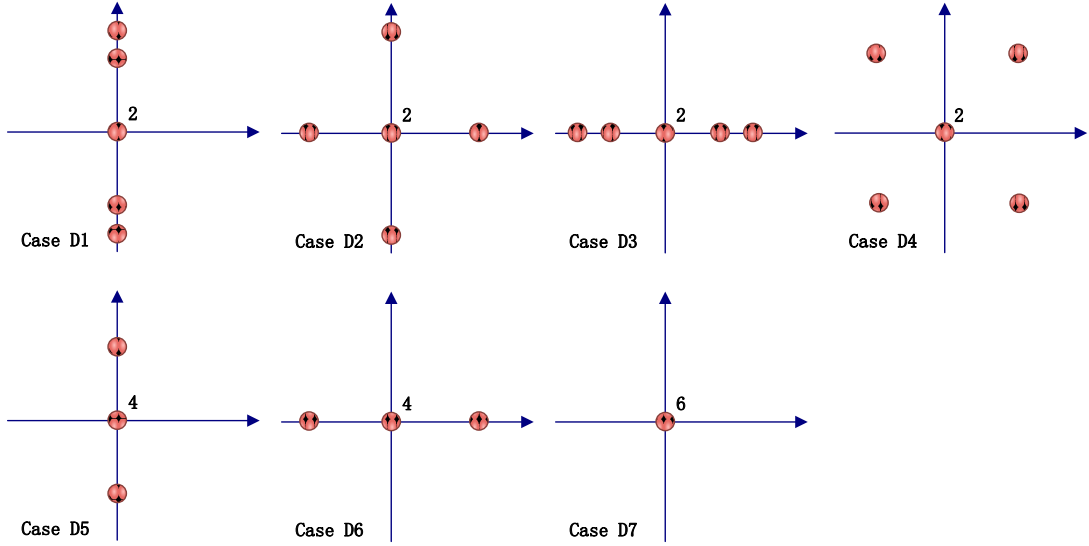

*Fig. 1. Topological Classification of degenerate equilibrium points*

The following theorem is about the eigenvalues and number of relative equilibria.

**Theorem 1.** Suppose that there are $N$ relative equilibria in the gravitational potential of an asteroid, denote $E_i$ as the $i$ th equilibrium point, $\lambda_j(E_i)$ as the $j$ th eigenvalue of the equilibrium point $E_i$, then we have $\sum_{i=1}^{N}\left[\operatorname{sgn}\prod_{j=1}^{6}\lambda_j(E_i)\right]=const$.

**Proof:**
The linearized equation relative to the relative equilibria can be given by

$$\mathbf{M}\ddot{\boldsymbol{\rho}}+\mathbf{G}\dot{\boldsymbol{\rho}}+\left(\nabla^2 V\right)\boldsymbol{\rho}=0. \quad (4)$$

Where $\boldsymbol{\rho}$ is the position relative to the equilibrium point, $\mathbf{M}$ is a $3\times 3$ unit matrix, $\mathbf{G}=\begin{pmatrix}0 & -2\omega & 0\\ 2\omega & 0 & 0\\ 0 & 0 & 0\end{pmatrix}$, $\omega$ is the norm of $\boldsymbol{\omega}$, and $\nabla^2 V$ is the Hessian matrix of the effective potential. Denote $\dot{\boldsymbol{\rho}}=\boldsymbol{\chi}$ and $\boldsymbol{\Lambda}=\begin{bmatrix}\boldsymbol{\chi}\\ \boldsymbol{\rho}\end{bmatrix}$, substituting into Eq. (4) yields

$$\dot{\mathbf{\Lambda}} = \mathbf{g}(\mathbf{\Lambda}) = \begin{pmatrix} \mathbf{0}_{3\times3} & \mathbf{I}_{3\times3} \\ -\mathbf{M}^{-1}(\nabla^2 V) & -\mathbf{M}^{-1}\mathbf{G} \end{pmatrix} \begin{bmatrix} \chi \\ \rho \end{bmatrix} = \mathbf{P}\mathbf{\Lambda}, \tag{5}$$

where

$$\mathbf{P} = \begin{pmatrix} \mathbf{0}_{3\times3} & \mathbf{I}_{3\times3} \\ -\mathbf{M}^{-1}(\nabla^2 V) & -\mathbf{M}^{-1}\mathbf{G} \end{pmatrix}. \tag{6}$$

Denote $f(\mathbf{r}) = \nabla V$, and thus we have $\dfrac{df}{d\mathbf{r}} = \nabla^2 V$ and $\det \mathbf{P} = \det(\nabla^2 V)$.

Denote $\Xi$ as the open set, using the topological degree theory in the function $f(\mathbf{r})$ yields

$$\sum_{i=1}^{N} \left[ \operatorname{sgn} \prod_{j=1}^{6} \lambda_i(E_k) \right] = \deg(f, \Xi, (0, 0, 0)) = const. \tag{7}$$

It can be rewritten as

$$\sum_{i=1}^{N} \left[ \operatorname{sgn} \prod_{j=1}^{6} \lambda_j(E_i) \right] = \sum_{i=1}^{N} \left[ \operatorname{sgn}(\det \mathbf{P}) \right] = \sum_{i=1}^{N} \left[ \operatorname{sgn}(\det(\nabla^2 V)) \right] = const. \tag{8}$$

□

The theorem gives a conserved quantity, which can restrict the number of non-degenerate equilibria in the gravitational potential of an asteroid. Thus, we have the following:

**Corollary 1.** The number of non-degenerate equilibria in the gravitational field of an asteroid varies in pairs.

**Corollary 2.** In regard to the degenerate equilibrium point in the gravitational field of an asteroid, it can follow one of the following four options:

1) disappear;

2) changes to $k(k \in Z, k > 0)$ degenerate equilibrium points;

3) changes to $2l(l \in Z, l > 0)$ non-degenerate equilibrium points;

4) changes to $k(k \in Z, k > 0)$ degenerate equilibrium points and $2l(l \in Z, l > 0)$ non-degenerate equilibrium points.

From the calculation of the number of equilibrium points, we have the following:

**Corollary 3.** The constant of the identity about the eigenvalues of the relative equilibria is 1, in other words $\sum_{i=1}^{N}\left[\text{sgn}\prod_{j=1}^{6}\lambda_j(E_i)\right]=1$. Thus, the number of non-degenerate equilibria in the gravitational field of an asteroid is an odd number, it may be 1, 3, 5, 7, 9, …, etc. If the parameter varies, the number of non-degenerate equilibria varies from one odd number to another odd number.

Elipe and Lara (2004) found four relative equilibria in the potential of a massive straight segment; Scheeres et al. (2004) found four relative equilibria around asteroid 25143 Itokawa; Scheeres et al. (2006) discovered four relative equilibria in the potential of the major body of binary asteroid 1999 KW4; Magri et al. (2011) discovered four relative equilibria around contact binary near-Earth Asteroid 1996 HW1; Yu and Baoyin (2012) discovered four relative equilibria around the major body of triple asteroid 216 Kleopatra; and Scheeres (2012) discovered six relative equilibria around asteroid 1580 Betulia and four relative equilibria around comet 67P/Churyumov-Gerasimenko. Jiang et al. (2014) found four relative equilibria for each of the three asteroids 1620 Geographos, 4769 Castalia, and 6489 Golevka. However, all of the above studies didn't calculate the relative equilibria inside the body of irregular minor celestial bodies. In Wang et al. (2014), relative equilibria of 23 irregular minor celestial bodies are calculated, including asteroids 4 Vesta, 216 Kleopatra, 243 Ida, 433 Eros, 951 Gaspra, 1620 Geographos, 1996 HW1, 1998 KY26, 2063 Bacchus, 2867 Steins, 4769 Castalia, 6489 Golevka, 25143 Itokawa, 52760, 101955 Bennu, satellites of planets J5 Amalthea, M1 Phobos, N8 Proteus, S9 Phoebe, S16

Prometheus, and comets 1P/Halley, 9P/Tempel1, as well as 103P/Hartley2. Relative equilibria of these 23 bodies have no zero eigenvalues, so these bodies have no degenerate equilibria. However, there exists a body with an odd number of equilibria with even number of them being degenerate or a body with an even number of equilibria with odd number of them being degenerate.

**3. Bifurcations and Collision of Relative Equilibria**

Denote $P$ as the parameter and $P_0$ as the parameter value at the bifurcation point. The conserved quantity from Theorem 1 can also restrict the kinds of bifurcations of relative equilibria in the gravitational potential of an asteroid when the parameter varies. Corollaries 4-10 give the possible bifurcations of relative equilibria in the gravitational potential of an asteroid.

**Corollary 4. Transcritical bifurcation**: When the parameter varies, the number of non-degenerate equilibria may vary from $2k+3\,(k \in Z, k \geq 0)$ to $2k+1\,(k \in Z, k \geq 0)$; meanwhile, two non-degenerate equilibria belonging to case 1 and case 2 collide and change to one degenerate equilibrium, which belongs to case D1, and then change to two non-degenerate equilibria belonging to case 2 and case 1,

$$\left.\begin{array}{l}\text{Case 1}\\ \text{Case 2}\end{array}\right\} \Rightarrow \text{Case D1} \Rightarrow \begin{cases}\text{Case 1}\\ \text{Case 2}\end{cases}.$$

**Corollary 5. Quasi-Transcritical bifurcation**: When the parameter varies, the number of non-degenerate equilibria may vary from $2k+3\,(k \in Z, k \geq 0)$ to $2k+1\,(k \in Z, k \geq 0)$; meanwhile, two non-degenerate equilibria belonging to case 2 and case 5 collide and change to one degenerate equilibrium, which belongs to case

D5, and then change to two non-degenerate equilibria belonging to case 5 and case 2,

$$\left.\begin{array}{l}\text{Case 2}\\ \text{Case 5}\end{array}\right\} \Rightarrow \text{Case D5} \Rightarrow \begin{cases}\text{Case 2}\\ \text{Case 5}\end{cases}.$$

**Corollary 6. Saddle-Node bifurcation**: When the parameter varies, the number of non-degenerate equilibria may vary from $2k+3\,(k \in Z, k \geq 0)$ to $2k+1\,(k \in Z, k \geq 0)$; meanwhile, two non-degenerate equilibria belonging to case 1 and case 2 collide and change to one degenerate equilibrium, which belongs to case D1, and then disappear, $\left.\begin{array}{l}\text{Case 1}\\ \text{Case 2}\end{array}\right\} \Rightarrow \text{Case D1} \Rightarrow \text{Disappear}$ or $\text{Disappear} \Rightarrow \text{Case D1} \Rightarrow \left.\begin{array}{l}\text{Case 1}\\ \text{Case 2}\end{array}\right\}$. In addition, the number of non-degenerate equilibria can also vary from $2k+1\,(k \in Z, k \geq 0)$ to $2k+3\,(k \in Z, k \geq 0)$; then a degenerate equilibrium appears and then separates into two non-degenerate equilibria, which belong to case 1 and case 2.

Suppose that when the parameter increases, the Saddle-Node bifurcation occurs. From Corollary 6, we have the collision of case 1 and case 2. From this situation, one can obtain that, when $P < P_0$, two non-degenerate equilibria belonging to case 1 and case 2 approach; when $P = P_0$, these two non-degenerate equilibria collided and become one degenerate equilibrium, which belongs to case D1; and when $P > P_0$, the degenerate equilibrium, which belongs to case D1, disappears.

**Corollary 7. Saddle-Saddle bifurcation**: When the parameter varies, the number of non-degenerate equilibria may vary from $2k+3\,(k \in Z, k \geq 0)$ to $2k+1\,(k \in Z, k \geq 0)$; meanwhile, two non-degenerate equilibria belonging to case 2 and case 5 collide and change to one degenerate equilibrium, which belongs to case

D5, and then disappear, $\left.\begin{array}{c}\text{Case 2}\\\text{Case 5}\end{array}\right\} \Rightarrow$ Case D5 $\Rightarrow$ Disappear or Disappear $\Rightarrow$ Case D5 $\Rightarrow \left.\begin{array}{c}\text{Case 2}\\\text{Case 5}\end{array}\right\}$. In addition, the number of non-degenerate equilibria can also vary from $2k+1(k \in Z, k \geq 0)$ to $2k+3(k \in Z, k \geq 0)$; then a degenerate equilibrium appears and then separates into two non-degenerate equilibria, which belong to case 2 and case 5, respectively.

**Corollary 8. Binary Saddle-Node bifurcation**: When the parameter varies, the number of non-degenerate equilibria may vary from $2k+3(k \in Z, k \geq 0)$ to $2k+1(k \in Z, k \geq 0)$; meanwhile, it can belong to one of the following two options:

1) Consider four non-degenerate equilibria, two of them belonging to case 1 and the other two belonging to case 2. One non-degenerate equilibrium belonging to case 1 and another non-degenerate equilibrium belonging to case 2 collide and change to one degenerate equilibrium, which belongs to case D1, and then change to one non-degenerate equilibrium, which belongs to case 1. Meanwhile, the other two equilibria, one non-degenerate equilibrium belonging to case 1 and another non-degenerate equilibrium belonging to case 2 collide and change to one degenerate equilibrium, which belongs to case D1, and then change to one non-degenerate equilibrium, which belongs to case 2.

$$\left.\begin{array}{l}\left.\begin{array}{c}\text{Case 1}\\\text{Case 2}\end{array}\right\} \Rightarrow \text{Case D1} \Rightarrow \text{Case 1}\\\left.\begin{array}{c}\text{Case 1}\\\text{Case 2}\end{array}\right\} \Rightarrow \text{Case D1} \Rightarrow \text{Case 2}\end{array}\right\}$$

2) Consider four non-degenerate equilibria, two of them belong to case 2 and the other two belong to case 5. One non-degenerate equilibrium belonging to case 2 and another non-degenerate equilibrium belonging to case 5 collide and change to one degenerate equilibrium, which belongs to case D5, and then change to one non-degenerate equilibrium, which belongs to case 2. Meanwhile, the other two equilibria, one non-degenerate equilibrium belonging to case 2 and another non-degenerate equilibrium belonging to case 5 collide and change to one degenerate equilibrium, which belongs to case D5, and then change to one non-degenerate equilibrium, which belongs to case 5.

$$\left.\begin{array}{l}\left.\begin{array}{l}\text{Case 2}\\ \text{Case 5}\end{array}\right\} \Rightarrow \text{Case D5} \Rightarrow \text{Case 2}\\ \left.\begin{array}{l}\text{Case 2}\\ \text{Case 5}\end{array}\right\} \Rightarrow \text{Case D5} \Rightarrow \text{Case 5}\end{array}\right\}$$

**Corollary 9. Supercritical Pitchfork bifurcation**: When the parameter varies, the number of non-degenerate equilibria may vary from $2k+3\,(k \in Z, k \geq 0)$ to $2k+1\,(k \in Z, k \geq 0)$; meanwhile, three non-degenerate equilibria, two of them belonging to case 1 and the other one belonging to case 2, collide at the bifurcation point and change to one non-degenerate equilibrium, which belongs to case 1. In addition, the number of non-degenerate equilibria can also vary from $2k+1\,(k \in Z, k \geq 0)$ to $2k+3\,(k \in Z, k \geq 0)$. One non-degenerate equilibrium, which belongs to case 1, becomes three non-degenerate equilibria, two of them belonging to case 1, and the other one belonging to case 2, $\left.\begin{array}{l}\text{Case 1}\\ \text{Case 2}\\ \text{Case 1}\end{array}\right\} \Leftrightarrow \text{Case 1}$.

**Corollary 10. Subcritical Pitchfork bifurcation**: When the parameter varies, the number of non-degenerate equilibria may vary from $2k+3\,(k \in Z, k \geq 0)$ to $2k+1\,(k \in Z, k \geq 0)$; meanwhile, it can belong to one of the following two options:

1) Three non-degenerate equilibria, two of them belonging to case 2 and the other one belonging to case 1, collide at the bifurcation point and change to one non-degenerate equilibrium, which belongs to case 2, $\left.\begin{array}{c}\text{Case 2}\\ \text{Case 1}\\ \text{Case 2}\end{array}\right\} \Leftrightarrow \text{Case 2}$;

2) Three non-degenerate equilibria, one belonging to each case 1, case 2, and case 5, collide at the bifurcation point and change to one non-degenerate equilibrium, which belongs to case 5, $\left.\begin{array}{c}\text{Case 1}\\ \text{Case 2}\\ \text{Case 5}\end{array}\right\} \Leftrightarrow \text{Case 5}$.

In addition, the number of non-degenerate equilibria can also vary from $2k+1 (k \in Z, k \geq 0)$ to $2k+3 (k \in Z, k \geq 0)$; meanwhile, one non-degenerate equilibrium, which belongs to case 1, becomes three non-degenerate equilibria, two of them belonging to case 2 and the other one belonging to case 1. Or one non-degenerate equilibrium, which belongs to case 4, becomes three non-degenerate equilibria one each belonging to case 1, case 2, and case 5.

Consider the triple asteroid system 216 Kleopatra. We now apply the result in order to calculate the bifurcation of relative equilibria when the parameter changes and see the collision and annihilation of the relative equilibria. 216 Kleopatra has overall dimensions of are $217 \times 94 \times 81$ km, rotation period 5.385 h (Ostro et al. 2000), and estimated bulk density 3.6 g·cm$^{-3}$ (Descamps et al. 2010). The gravitational potential and shape model of 216 Kleopatra is computed by the polyhedral method (Werner and Scheeres 1997) using data from radar observations (Neese 2004), which includes 2048 vertices and 4096 faces. Asteroid 216 Kleopatra has seven equilibria (Jiang et al. 2014; Wang et al. 2014; Hirabayashi and Scheeres 2014), and all of them are non-degenerate. Four of them are outside the body of 216 Kleopatra, and all of them are unstable (Yu and Baoyin 2012; Jiang et al. 2014; Hirabayashi and Scheeres 2014); while the other three are inside the body, and two of them are stable and the other one is unstable (Wang et al. 2014). Figure 2 shows the effective potential of asteroid 216 Kleopatra in the equatorial plane. From Figure 2, it can clearly be see that there are 7 critical points of effective potential in the equatorial plane.

Considering the rotation speed $\omega$ as the parameter, the number of relative equilibria of asteroid 216 Kleopatra change if the parameter changes. If $1.0\omega_0 \leq \omega \leq 4.270773\omega_0$, where $\omega_0 = \frac{2\pi}{5.385}\text{h}^{-1}$, then the rotation speed is magnified. The result shows that $\omega = 1.944586\omega_0$, $\omega = 2.03694\omega_0$, and $\omega = 4.270773\omega_0$ are bifurcation points; and two equilibria collide and annihilate each other at the bifurcation points $\omega = 1.944586\omega_0$, $\omega = 2.03694\omega_0$, and $\omega = 4.270773\omega_0$. Figure 3 shows the position of equilibrium points when the parameter changes from $1.0\omega_0$ to $\omega = 4.270773\omega_0$. If the parameter equals a value bigger than $4.270773\omega_0$, then the number, topological classification, stability, and positive index of inertia are the same as when $\omega = 4.270773\omega_0$. Table A1 in the Appendix shows the position of relative equilibria when the parameter changes; while Table A2 in the Appendix shows the topological classification and stability of relative equilibria when the parameter changes.

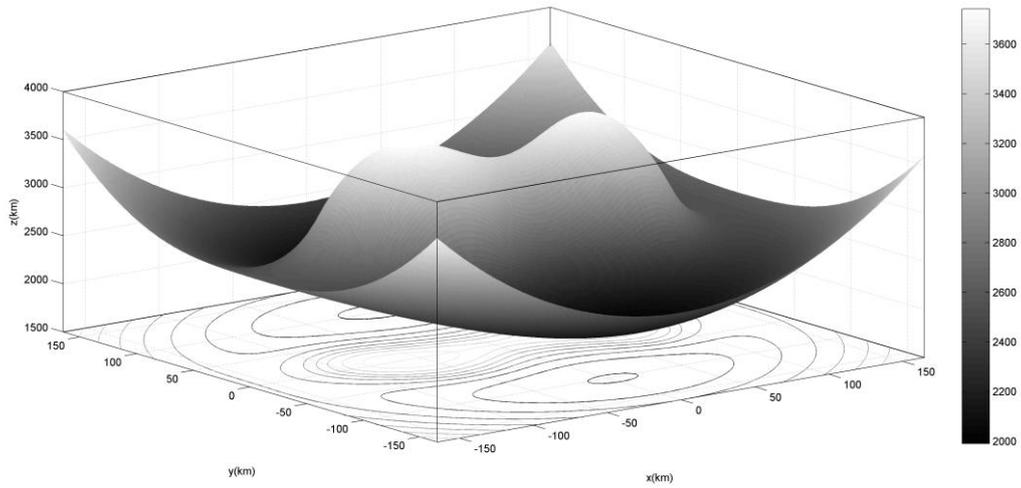

Figure 2. The effective potential of asteroid 216 Kleopatra in the equatorial plane, the unit is m$^2$ s$^{-2}$.

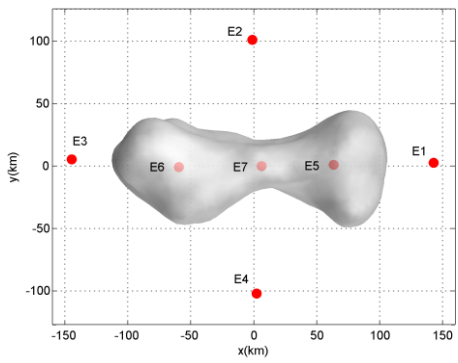

(a)

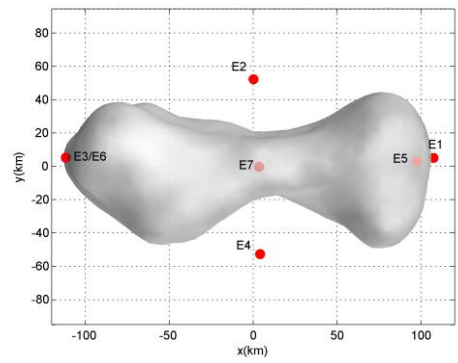

(b)

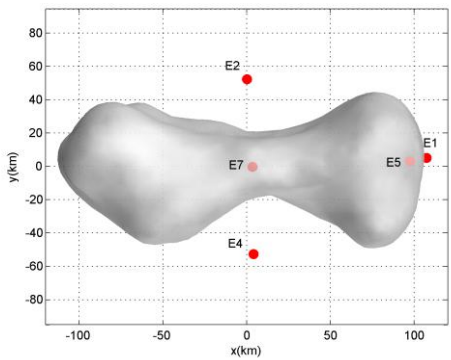

(c)

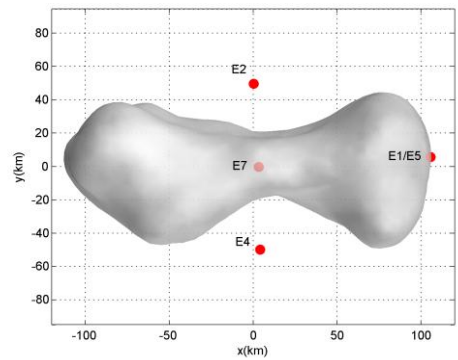

(d)

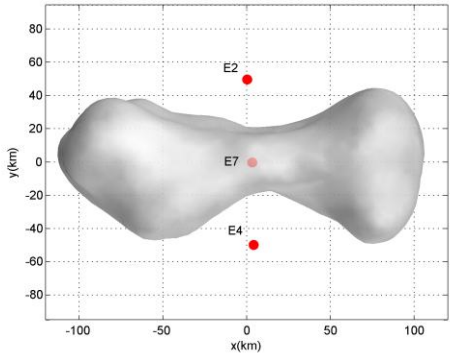
(e)

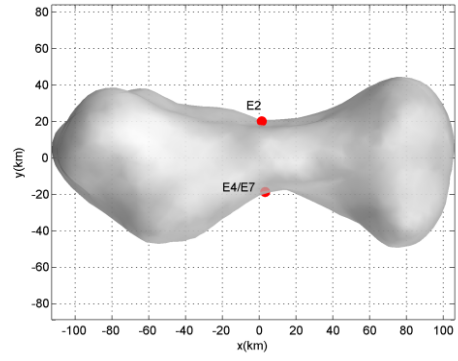
(f)

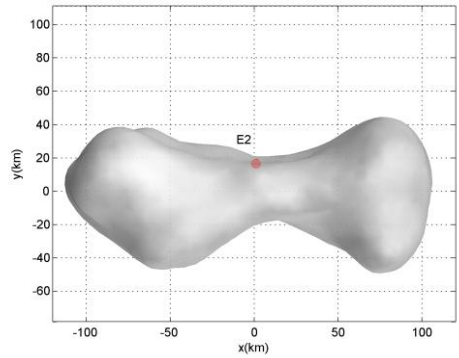
(g)

Figure 3. The location of relative equilibria of asteroid 216 Kleopatra: (a) $\omega=1.0\omega_0$; (b) $\omega=1.944586\omega_0$; (c) $\omega=1.944587\omega_0$; (d) $\omega=2.03694\omega_0$; (e) $\omega=2.03695\omega_0$; (f) $\omega=4.270772\omega_0$; and (g) $\omega=4.270773\omega_0$.

In the beginning, the parameter is $\omega=1.0\omega_0$ and there are 7 equilibrium points. If the parameter varies from $\omega=1.0\omega_0$ to $\omega=1.944586\omega_0$, the positions of the equilibrium points vary and equilibrium points E3 and E6 collide and become a degenerate equilibrium point, the topological case of E3/E6 becoming D1; while the parameter magnifies from $\omega=1.944586\omega_0$ to $\omega=1.944587\omega_0$, the degenerate equilibrium point E3/E6 vanishes. There are now only 5 equilibrium points left, which are E1, E2, E4, E5, and E7. The bifurcation in this situation is a saddle-node bifurcation. Let the parameter change from $\omega=1.944587\omega_0$ to $\omega=2.03694\omega_0$, the positions of the equilibrium points vary and equilibrium points E1 and E5 collide and become a degenerate equilibrium point, the topological case of E1/E5 becoming D1; while the parameter magnifies from $\omega = 2.03694\omega_0$ to $\omega = 2.03695\omega_0$, the degenerate equilibrium point E1/E5 vanishes. There are now only 3 equilibrium points left, which are E2, E4, and E7. The bifurcation in this situation is also a saddle-node bifurcation. Let the parameter change from $\omega = 2.03695\omega_0$ to $\omega = 4.270772\omega_0$, the positions of the equilibrium points vary and equilibrium points E4 and E7 collide and become a degenerate equilibrium point, the topological case of E4/E7 becoming D5; while the parameter magnifies from $\omega = 4.270772\omega_0$ to $\omega = 4.270773\omega_0$, the degenerate equilibrium point E4/E7 vanishes. There is now only 1 equilibrium point left, which is E2. The bifurcation in this situation is a saddle-saddle bifurcation. From Figure 3g, it can be seen that when $\omega = 4.270773\omega_0$, there is only equilibrium point E2 left.

From Tables A1 and A2 in the Appendix, one can see that, while the parameter varies, all the equilibria's positions vary. When the parameter varies, there only 3 kinds of non-degenerate equilibria exist, which are cases 1, 2, and 5; and only 2 kinds of degenerate equilibria exist, which are cases D1 and D5. From Figure 3a and Table A1, it can be seen that, when $\omega = 1.0\omega_0$, there are 4 and 3 non-degenerate equilibrium points outside and inside the body of 216 Kleopatra, respectively. The positions of the equilibria has some symmetrical characteristics, the equilibrium points, which are near the x-axis/y-axis and opposite to the barycentre, have the same topological case, including E1 and E3, E2 and E4, and E5 and E6. The symmetrical characteristics are retained when $1.0\omega_0 \leq \omega < 1.944586\omega_0$, and is lost when $1.944586\omega_0 \leq \omega \leq 2.03694\omega_0$. When $1.944586\omega_0 \leq \omega < 2.03695\omega_0$, only equilibrium points E2 and E4, which are near the y-axis and opposite to the barycentre and have the same topological case, retain these symmetrical characteristics. When $\omega \geq 2.03695\omega_0$, the symmetrical characteristic are no longer retained.

The linear stability; positive, definite, or not; of the Hessian matrix of the effective potential and the positive/negative index of inertia are also calculated and presented in Table A2 of the Appendix. When $\omega = 1.0\omega_0$, equilibrium points E5 and E6 are linear stable, the Hessian matrix of the effective potential at E5 and E6 is positive definite, and the positive index of inertia of the Hessian matrix is 3. When $\omega = 4.270772\omega_0$, there are only 2 equilibrium points, the degenerate equilibrium point E4/E7 and the non-degenerate equilibrium point E2. The equilibrium point E2 is linear and stable, but the Hessian matrix of the effective potential at E2 is non-positive definite, and the positive and negative index of inertia of the Hessian matrix at E2 are 1 and 2, respectively. This implies that the equilibrium point E2 is unstable, although it is linearly stable. When $\omega = 4.270773\omega_0$, only 1 equilibrium point exists, which is the non-degenerate equilibrium point E2. Equilibrium point E2 is also linear stable, and the Hessian matrix of the effective potential at E2 is non-positive definite. Thus, the equilibrium point E2 is unstable and linearly stable.

## 4. Conclusions

This paper analyzes the collision and annihilation of relative equilibria in the gravitational potential of asteroids when the parameter, rotation speed, varies. An identical equation, which is about the relationship between the eigenvalues and the number of relative equilibria, has been presented. It is impossible for the number of non-degenerate equilibria to be an even number. The conserved quantity can restrict the kinds of bifurcations of relative equilibria. Several kinds of bifurcations are forecasted. When the rotation speed is taken as a parameter, the number of relative equilibria varies if the rotation speed varies and equilibrium points are likely to collide and annihilate each other. The saddle-node bifurcation and the saddle-saddle bifurcation are discovered when the rotation period of asteroid 216 Kleopatra varies.

**Appendix 1**

Table A1. The position of relative equilibria of asteroid 216 Kleopatra when the parameter changes

$\omega = 1.0\omega_0$

| Equilibrium Points | x (km) | y (km) | z (km) |
|---|---|---|---|
| E1 | 142.8529 | 2.44128 | 1.18154 |
| E2 | -1.16386 | 100.740 | -0.545911 |
| E3 | -144.684 | 5.18876 | -0.272475 |
| E4 | 2.22996 | -102.103 | 0.271873 |
| E5 | 63.4441 | 0.827510 | -0.694539 |
| E6 | -59.5426 | -0.969171 | -0.191990 |
| E7 | 6.21921 | -0.198684 | -0.308407 |

$\omega = 1.944586\omega_0$

| Equilibrium Points | x (km) | y (km) | z (km) |
|---|---|---|---|
| E1 | 107.364 | 4.94478 | 3.92404 |
| E2 | 0.205470 | 52.1013 | -1.36327 |
| E3/E6 | -111.610 | 5.10818 | -3.95857 |

| Equilibrium Points | x (km) | y (km) | z (km) |
| --- | --- | --- | --- |
| E4 | 4.11958 | -52.7590 | 0.601498 |
| E5 | 97.3692 | 2.95336 | 2.67332 |
| E7 | 3.52509 | -0.376938 | -0.209140 |

$$\omega = 1.944587\omega_0$$

| Equilibrium Points | x (km) | y (km) | z (km) |
| --- | --- | --- | --- |
| E1 | 107.364 | 4.94478 | 3.92404 |
| E2 | 0.205470 | 52.1013 | -1.36328 |
| E4 | 4.11959 | -52.7591 | 0.601498 |
| E5 | 97.3692 | 2.95337 | 2.67332 |
| E7 | 3.52509 | -0.376939 | -0.209140 |

$$\omega = 2.03694\omega_0$$

| Equilibrium Points | x (km) | y (km) | z (km) |
| --- | --- | --- | --- |
| E1/E5 | 105.630 | 5.54863 | 4.40158 |
| E2 | 0.310281 | 49.4763 | -1.43647 |
| E4 | 4.17061 | -49.9763 | 0.638675 |
| E7 | 3.33339 | -0.398195 | -0.202037 |

$$\omega = 2.03695\omega_0$$

| Equilibrium Points | x (km) | y (km) | z (km) |
| --- | --- | --- | --- |
| E2 | 0.310282 | 49.4763 | -1.43648 |
| E4 | 4.17062 | -49.9764 | 0.638676 |
| E7 | 3.33339 | -0.398196 | -0.202037 |

$$\omega = 4.270772\omega_0$$

| Equilibrium Points | x (km) | y (km) | z (km) |
| --- | --- | --- | --- |
| E2 | 1.09452 | 16.4117 | -2.66997 |
| E4/E7 | 3.11753 | -18.3502 | 2.60562 |

$$\omega = 4.270773\omega_0$$

| Equilibrium Points | x (km) | y (km) | z (km) |
| --- | --- | --- | --- |
| E2 | 1.09452 | 16.4116 | -2.66995 |

Table A2. The topological classification and stability of the relative equilibria of asteroid 216 Kleopatra when the parameter changes. LS: linearly stable; U: unstable; D: degenerate; P: positive definite; N: non-positive definite; Index of inertia: positive/ negative index of inertia

$$\omega = 1.0\omega_0$$

| Equilibrium Points | Case | Stability | $\nabla^2 V$ | Index of Inertia |
|---|---|---|---|---|
| E1 | 2 | U | N | 2/1 |
| E2 | 5 | U | N | 1/2 |
| E3 | 2 | U | N | 2/1 |
| E4 | 5 | U | N | 1/2 |
| E5 | 1 | LS | P | 3/0 |
| E6 | 1 | LS | P | 3/0 |
| E7 | 2 | U | N | 2/1 |

$$\omega = 1.944586\omega_0$$

| Equilibrium Points | Case | Stability | $\nabla^2 V$ | Index of Inertia |
|---|---|---|---|---|
| E1 | 2 | U | N | 2/1 |
| E2 | 5 | U | N | 1/2 |
| E3/E6 | D1 | D | N | 2/0 |
| E4 | 5 | U | N | 1/2 |
| E5 | 1 | LS | P | 3/0 |
| E7 | 2 | U | N | 2/1 |

$$\omega = 1.944587\omega_0$$

| Equilibrium Points | Case | Stability | $\nabla^2 V$ | Index of Inertia |
|---|---|---|---|---|
| E1 | 2 | U | N | 2/1 |
| E2 | 5 | U | N | 1/2 |
| E4 | 5 | U | N | 1/2 |
| E5 | 1 | LS | P | 3/0 |
| E7 | 2 | U | N | 2/1 |

$$\omega = 2.03694\omega_0$$

| Equilibrium Points | Case | Stability | $\nabla^2 V$ | Index of Inertia |
|---|---|---|---|---|
| E1/E5 | D1 | D | N | 2/0 |
| E2 | 5 | U | N | 1/2 |
| E4 | 5 | U | N | 1/2 |
| E7 | 2 | U | N | 2/1 |

$$\omega = 2.03695\omega_0$$

| Equilibrium Points | Case | Stability | $\nabla^2 V$ | Index of Inertia |
|---|---|---|---|---|
| E2 | 5 | U | N | 1/2 |
| E4 | 5 | U | N | 1/2 |
| E7 | 2 | U | N | 2/1 |

$$\omega = 4.270772\omega_0$$

| Equilibrium Points | Case | Stability | $\nabla^2 V$ | Index of Inertia |
|---|---|---|---|---|
| E2 | 1 | LS | N | 1/2 |
| E4/E7 | D5 | D | N | 2/0 |

$$\omega = 4.270773\omega_0$$

| Equilibrium Points | Case | Stability | $\nabla^2 V$ | Index of Inertia |
|---|---|---|---|---|
| E2 | 1 | LS | N | 1/2 |

**Acknowledgements**

This research is supported by the National Basic Research Program of China (973 Program, 2012CB720000), the State Key Laboratory Foundation of Astronautic Dynamics (No. 2013ADL-DW02) and the National Natural Science Foundation of China (No. 11372150).

# References

Asphaug, S. J., Ostro, R. S., Hudson, D. J.: Disruption of kilometer-sized asteroids by energetic collisions. Nature 393, 437-440 (1998)

Campins, H., Hargrove, K., Pinilla-Alonso, N., et al.: Water ice and organics on the surface of the asteroid 24 Themis. Nature 2010, 464(7293): 1320-1321.

Chesley, S. R., Ostro, S. J., Vokrouhlický, D., et al.: Direct detection of the Yarkovsky effect by radar ranging to asteroid 6489 Golevka. Science 2003, 302(5651): 1739-1742.

Descamps, P., Marchis, F., Berthier, J. et al.: Triplicity and physical characteristics of Asteroid (216) Kleopatra. Icarus 211(2), 1022-1033(2011)

Ďurech, J., Vokrouhlický, D., Kaasalainen, M., et al.: Detection of the YORP effect in asteroid (1620) Geographos. Astron. Astrophys. 2008, 489(2): L25-L28.

Elipe, A., Lara, M.: A simple model for the chaotic motion around (433) Eros. Adv. Astron. Sci. 116, 1-15(2004)

Hartmann, W. K.: The shape of Kleopatra. Science 288 (5467), 820-821 (2000)

Hirabayashi, M., Scheeres, D. J.: Analysis of asteroid (216) Kleopatra using dynamical and structural constraints. Astrophys. J. 780(2): 160. (2014)

Jiang, Y., Baoyin, H., Li, J., Li, H.: Orbits and manifolds near the equilibrium points around a rotating asteroid. Astrophys. Space Sci. 349(1), 83-106 (2014)

Jiang, Y., Baoyin, H.: Orbital mechanics near a rotating asteroid. J. Astrophys. Astron. 35(1): 17-38(2014)

Jiang, Y.: Equilibrium points and periodic orbits in the vicinity of asteroids with an application to 216 Kleopatra. Earth, Moon and Planets. 115(1-4) 31-44 (2015)

Jiang, Y., Yu, Y., Baoyin, H.: Topological classifications and bifurcations of periodic orbits in the potential field of highly irregular-shaped celestial bodies. Nonlinear Dynam. 81(1-2), 119-140 (2015)

Li, X., Qiao, D., Cui, P.: The equilibria and periodic orbits around a dumbbell-shaped body. Astrophys. Space Sci. 348(2): 417-426 (2013)

Liu, X., Baoyin, H., Ma, X.: Equilibria, periodic orbits around equilibria, and heteroclinic connections in the gravity field of a rotating homogeneous cube. Astrophys. Space Sci. 333(2), 409-418 (2011)

Lupishko, D., Tielieusova, I.: Influence of the YORP effect on rotation rates of near‐Earth asteroids. Meteori. Planet. Sci. 49(1), 80-85(2014)

Magri, C., Howell, E. S., Nolan, M. C.et al.: Radar and photometric observations and shape modeling of contact binary near-Earth Asteroid (8567) 1996 HW1. Icarus 214(1), 210-227(2011)

Michel, P., Benz, W., Tanga, P., et al.: Collisions and gravitational reaccumulation: Forming asteroid families and satellites. Science 294(5547), 1696-1700(2001)

Nesvorný, D., Bottke, Jr. W. F., Dones, L., et al.: The recent breakup of an asteroid in the main-belt region. Nature 417(6890), 720-771 (2002)

Ostro, S. J., Hudson, R. S., Nolan, M. C. et al.: Radar observations of asteroid 216 Kleopatra. Science 288(5467), 836-839 (2000)

Palacián, J. F., Yanguas, P., Gutiérrez-Romero, S.: Approximating the invariant sets of a finite straight segment near its collinear equilibria. SIAM J. Appl. Dyn. Syst. 5(1), 12-29(2006)

Richardson, D. C., Elankumaran, P., Sanderson, R. E.: Numerical experiments with rubble piles: equilibrium shapes and spins. Icarus 173(2),349-361(2005)

Richardson, J. E., Melosh, H. J., Greenberg, R.: Impact-induced seismic activity on Asteroid 433 Eros: A surface modification process. Science 306(5701),1526-1529 (2004)

Romanov, V. A., Doedel, E. J.: Periodic orbits associated with the libration points of the homogeneous rotating gravitating triaxial ellipsoid. Int. J. Bifurcat. Chaos 22(10). 1230035(2012)

Rubincam, D. P.: Radiative spin-up and spin-down of small asteroids. Icarus 148(1), 2-11(2000)

Scheeres, D. J.: Orbital mechanics about small bodies. Acta Astronaut. **7**, 21-14 (2012)

Scheeres, D. J., Broschart, S., Ostro, S. J., Benner, L.: The dynamical environment about Asteroid 25143 Itokawa: target of the Hayabusa Mission. Proceedings of the AIAA/AAS Astrodynamics Specialist Conference and Exhibit. (2004)

Scheeres, D. J., Fahnestock, E. G., Ostro, S. J. et al.: Dynamical configuration of binary near-Earth asteroid (66391) 1999 KW4. Science 314(5803), 1280-1283 (2006)

Scheeres, D. J., Ostro, S. J., Hudson, R. S., et al.: Orbits close to asteroid 4769 Castalia, Icarus 121, 67-87 (1996)

Taylor, P. A., Margot, J. L., Vokrouhlický, D., et al.: Spin rate of asteroid (54509) 2000 PH5 increasing due to the YORP effect. Science 316(5822), 274-277(2007)

Walsh, K. J., Richardson, D. C., Michel, P.: Rotational breakup as the origin of small binary asteroids. Nature 454(7201),188-191(2008)

Walsh, K. J., Richardson, D. C., Michel, P.: Spin-up of rubble-pile asteroids: Disruption, satellite formation, and equilibrium shapes. Icarus 220(2), 514-529 ( 2012)

Wang, X., Jiang, Y., Gong, S.: Analysis of the potential field and equilibrium points of irregular-shaped minor celestial bodies. Astrophys. Space Sci. 353(1), 105-121(2014)

Werner, R. A., Scheeres, D. J.: Exterior gravitation of a polyhedron derived and compared with harmonic and mascon gravitation representations of asteroid 4769 Castalia. Celest. Mech. Dyn. Astron. 65(3), 313-344 (1997)

Yu, Y., Baoyin, H.: Orbital dynamics in the vicinity of asteroid 216 Kleopatra. Astron. J. 143(3), 62-70(2012)